# Energy-Efficient Downlink Power Control in mmWave Cell-Free and User-Centric Massive MIMO

Mario Alonzo[1], Stefano Buzzi[1], and Alessio Zappone[2,1]
[1]Department of Electrical and Information Engineering, University of Cassino and Lazio Meridionale, Cassino, Italy
[2] Large Systems and Networks Group, CentraleSupélec, Université Paris-Saclay, Gif-sur-Yvette, France.

*Abstract*—This paper considers cell-free and user-centric approaches for coverage improvement in wireless cellular systems operating at millimeter wave frequencies, and proposes downlink power control algorithms aimed at maximizing the global energy efficiency. To tackle the non-convexity of the problems, an interaction between sequential and alternating optimization is considered. The use of hybrid analog/digital beamformers is also taken into account. The numerical results show the benefits obtained from the power control algorithm, as well as that the user-centric approach generally outperforms the cell-free one.

## I. INTRODUCTION

Future fifth-generation (5G) wireless systems will heavily rely on the combined use of large-scale antenna arrays, a.k.a. massive MIMO, and of carrier frequencies above 10GHz, the so called mmWave frequencies [1]. For conventional sub-6 GHz frequencies, a new communications architecture, named "Cell-Free" (CF) massive MIMO, was introduced in [2], in order to alleviate the cell-edge problem and thus increase the system performance of unlucky users that happen to be located very far from their serving access point (AP). In the CF architecture, instead of having few base stations with massive antenna arrays, a very large number of simple APs randomly and densely deployed – and connected to a central processing unit (CPU) – serve a much smaller number of mobile stations (MSs). In [3], [4], the CF architecture has been generalized to the case in which both the APs and the MSs are equipped with multiple antennas and, most importantly, a user-centric (UC) variant of the CF approach is considered, wherein each APs, instead of serving all the MSs in the considered area, just serves the ones that he receives best. The results in [3], [4] show that the UC approach provides savings on the required backhaul capacity and, also, provides better data-rates than the CF approach to the vast majority of the users.

The recent paper [5] is the first to consider, to the best of authors' knowledge, the CF and UC approaches for a system operating at mmWave frequencies. Motivated by the fact that bit-per-Joule energy efficiency is regarded as a key requirement of future 5G networks [6], this paper generalizes the results of [5] by considering downlink power control rules aimed at maximizing the system global energy efficiency (GEE) in a wireless system using mmWave frequencies, and taking into account also the presence of hybrid analog/digital beamforming; see also [7], [8] on energy-efficient radio resource allocation for 5G systems. The proposed algorithm permits, as a by-product, to derive also the power allocation aimed at achievable spectral efficiency (ASE) maximization.

## II. SYSTEM MODEL

We consider an area where $K$ MSs and $M$ APs are randomly located. The APs are connected by means of a backhaul network to a CPU wherein data-decoding is performed. Communications take place on the same frequency band; downlink and uplink are separated through TDD. The communication protocol is made of three different phases: uplink training, downlink data transmission and uplink data transmission (not considered in this paper). While in the CF approach all the APs simultaneously serve all the MSs (a fully-cooperative scenario), in the UC approach each AP serves a pre-determined number of MSs, say $N$, and in particular the ones that it receives best.

### A. Channel model

We assume that each AP (MS) is equipped with a uniform linear array (ULA) with $N_{AP}$ ($N_{MS}$) elements. The ($N_{AP} \times N_{MS}$)-dimensional matrix $\mathbf{H}_{k,m}$ denotes the channel matrix between the $k$-th user and the $m$-th AP. According to the widely used clustered channel model for mmWave frequencies (see [9] and references therein), $\mathbf{H}_{k,m}$ can be expressed as

$$\mathbf{H}_{k,m} = \gamma \sum_{i=1}^{N_{cl}} \sum_{l=1}^{N_{ray}} \alpha_{i,l} \sqrt{L(r_{i,l})} \mathbf{a}_{AP}(\theta^{AP}_{i,l,k,m}) \mathbf{a}^H_{MS}(\theta^{MS}_{i,l,k,m}) + \mathbf{H}_{LOS}, \quad (1)$$

where $N_{cl}$ is the number of clusters, $N_{ray}$ is the number of the rays that we consider for each cluster, $\gamma$ is a normalization factor defined as $\sqrt{\dfrac{N_{AP} N_{MS}}{N_{cl} N_{ray}}}$, $\mathbf{H}_{LOS}$ is the line-of-sight (LOS) component, $\alpha_{i,l}$ is the complex path gain distributed as $\mathcal{CN}(0, \sigma^2)$ where $\sigma^2 = 1$, $L(r_{i,l})$ is the attenuation related to the path $(i, j)$, $\mathbf{a}_{AP}$ and $\mathbf{a}_{MS}$ are the ULA array responses at the $m$-th AP and at the $k$-th MS, respectively, and they depend on the angles of arrival and departure, $\theta^{AP}_{i,l,k,m}$ and $\theta^{MS}_{i,l,k,m}$, relative to the $(i, l)$-th path of the channel between

the $k$-th MS and the $m$-th AP. The $\mathbf{H}_{LOS}$ in (1) is written as[1]

$$\mathbf{H}_{LOS} = I(d)\sqrt{N_{AP}N_{MS}}e^{j\eta}\sqrt{L(d)}\mathbf{a}_{AP}(\theta_{LOS}^{AP})\mathbf{a}_{MS}^H(\theta_{LOS}^{MS}). \quad (2)$$

In the above equation, $\eta \sim \mathcal{U}(0, 2\pi)$, $I(d)$ is a 0-1 random variate indicating if a LOS link exists between the transmitter and the receiver, and $d$ is the link length. Due to lack of space, we refer the reader to references [5], [9] for additional details on the channel model.

## III. THE COMMUNICATION PROTOCOL

In the following, we assume that each MS employs a very simple 0-1 beamforming structure; in particular, the $(N_{MS} \times P)$-dimensional beamformer ($P$ is the multiplexing order) used at the $k$-th MS is denoted by $\mathbf{L}_k$ and is defined as $\mathbf{L}_k = \mathbf{I}_P \otimes \mathbf{1}_{N_{MS}/P}$, with $\otimes$ denoting Kronecker product and $\mathbf{1}_{N_{MS}/P}$ an all-1 vector of length $N_{MS}/P$. Otherwise stated, we assume that the MS receive antennas are divided in $P$ disjoint groups of $N_{MS}/P$ elements, and the data collected at the antennas of each group are simply summed together. We describe now the phases of the communication protocol.

### A. Uplink training

Let $\tau_c$ be the length of the channel coherence time and $\tau_p$ be the length of uplink training phase, both in discrete time samples. Of course we must have $\tau_p < \tau_c$. We define by $\mathbf{\Phi}_k \in \mathcal{C}^{P \times \tau_p}$ the matrix containing on its rows the pilot sequences sent by the $k$-th MS. We assume that $\mathbf{\Phi}_k \mathbf{\Phi}_k^H = \mathbf{I}_P$, i.e. the rows of $\mathbf{\Phi}_k$ are orthogonal, but no orthogonality is required for the pilot sequences assigned to other MSs[2]. The received signal at the $m$-th AP in the $\tau_p$ signaling intervals devoted to uplink training can be cast in the following $N_{AP} \times \tau_p$-dimensional matrix $\mathbf{Y}_m$:

$$\mathbf{Y}_m = \sum_{k=1}^{K} \sqrt{p_k} \mathbf{H}_{k,m} \mathbf{L}_k \mathbf{\Phi}_k + \mathbf{W}_m, \quad (3)$$

where $\mathbf{W}_m$ is the matrix of thermal noise samples, whose entries are assumed to be i.i.d. $\mathcal{CN}(0, \sigma_w^2)$ RVs. Letting now $\mathbf{S}_{k,m} = \mathbf{H}_{k,m} \mathbf{L}_k$, at the $m$-th AP, an estimate for the quantities $\{\mathbf{S}_{k,m}\}_{k=1}^{K}$ can be obtained as follows:

$$\widehat{\mathbf{S}}_{k,m} = \frac{1}{\sqrt{p_k}} \mathbf{Y}_m \mathbf{\Phi}_k^H = \mathbf{H}_{k,m}\mathbf{L}_k + \sum_{l=1, l \neq k}^{K} \sqrt{\frac{p_l}{p_k}} \mathbf{H}_{l,m} \mathbf{L}_l \mathbf{\Phi}_l \mathbf{\Phi}_k^H + \frac{1}{\sqrt{p_k}} \mathbf{W}_m \mathbf{\Phi}_k^H. \quad (4)$$

The estimation must be performed in all APs, i.e. for all $m = 1, \ldots, M$ and for all $k = 1, \ldots, K$. Of course, more sophisticated channel estimation schemes can be applied but here we are targeting an extremely simple system processing.

[1]For the ease of notation we omit the pedices $k, m$.
[2]Of course, when $KP \leq \tau_p$ it would be possible to assign to all the MSs mutually orthogonal pilot sequences. In this paper, however, we assume that the pilot sequences are binary random sequences, and we just require that each matrix $\mathbf{\Phi}_k$ has orthogonal rows.

### B. Downlink data transmission

After the first phase, the generic $m$-th AP has an estimate of the quantities $\mathbf{S}_{k,m}$, for all $k = 1, \ldots, K$. In order to transmit data on the downlink, a zero-forcing precoder is considered. First of all, we can define the following matrix: $\mathcal{G}_{k,m} = [\mathbf{S}_{1,1} \ldots \mathbf{S}_{K,M}] \forall k = 1, \ldots, K, m = 1, \ldots, M$, and then we define the precoding matrix as follows:

$$\mathbf{Q}_{k,m} = (\mathcal{G}_{k,m}\mathcal{G}_{k,m}^H)^{-1} \mathbf{S}_{k,m}. \quad (5)$$

Then, each precoding matrix has to be normalized as follows:

$$\mathbf{Q}_{k,m} = \frac{\mathbf{Q}_{k,m}}{\sqrt{tr(\mathbf{Q}_{k,m}\mathbf{Q}_{k,m}^H)}}, \forall k = 1, \ldots, K, m = 1, \ldots, M \quad (6)$$

The previously described beamforming matrix is a fully-digital (FD) one. We will also consider lower-complexity hybrid analog/digital approximations of the FD beamformer, by exploiting the *Block Coordinate Descent algorithm* [10].

*1) The UC approach:* In this case the generic $m$-th AP serves the $N$ MSs whose channels have the largest Frobenious norms. We denote by $\mathcal{K}(m)$ the set of MSs served by the $m$-th AP. Given the sets $\mathcal{K}(m)$, for all $m = 1, \ldots, M$, we can define the set $\mathcal{M}(k)$ of the APs that communicate with the $k$-th user:

$$\mathcal{M}(k) = \{m : k \in \mathcal{K}(m)\}. \quad (7)$$

In this case the transmitted signal from the $m$-th AP is written as:

$$\mathbf{s}_m^{UC}(n) = \sum_{k \in \mathcal{K}(m)} \sqrt{\eta_{m,k}^{DL,UC}} \mathbf{Q}_{k,m}^{DL} \mathbf{x}_k^{DL}(n), \quad (8)$$

where $\mathbf{x}_k^{DL}(n)$ is the data symbol intended for the $k$-th MS, and $\eta_{m,k}^{DL,UC}$ is a scalar coefficient ruling the transmitted power used at the $m$-th AP for the $k$-th MS data symbol. The received signal at the $k$-th MS is expressed as:

$$\mathbf{r}_k^{UC}(n) = \sum_{m=1}^{M} \mathbf{H}_{k,m}^H \mathbf{s}_m^{UC}(n) + \mathbf{z}_k(n) =$$
$$= \sum_{m \in \mathcal{M}(k)} \sqrt{\eta_{m,k}^{DL,UC}} \mathbf{H}_{k,m}^H \mathbf{Q}_{k,m}^{DL} \mathbf{x}_k^{DL}(n) +$$
$$+ \sum_{l=1, l \neq k}^{K} \sum_{m \in \mathcal{M}(l)} \sqrt{\eta_{m,l}^{DL,UC}} \mathbf{H}_{k,m}^H \mathbf{Q}_{l,m}^{DL} \mathbf{x}_l^{DL}(n) + \mathbf{z}_k(n), \quad (9)$$

where $\mathbf{z}_k(n)$ is the $\mathcal{CN}(0, \sigma_z^2)$ additive thermal noise. It is thus possible to obtain a soft estimate of the data symbol $\mathbf{x}_k^{DL}(n)$ at the $k$-th MS as $\hat{\mathbf{x}}_k^{DL,UC}(n) = \mathbf{L}_k^H \mathbf{r}_k^{UC}(n)$.

*2) The CF approach:* In this case, all the APs communicate with all the MSs, so the CF approach can be seen as a special case of the UC approach by considering $\mathcal{K}(m) = \{1, 2, \ldots, K\}, \forall m = 1, 2, \ldots, M$.

## C. Downlink ASE

Now, it is possible to write the ASE for the $k$-th user in the UC approach as follows[3]

$$\mathcal{R}_k = B \log_2 \det[\mathbf{I} + \mathbf{R}_k^{-1} \mathbf{A}_{k,k} \mathbf{A}_{k,k}^H] \quad (10)$$

where $\mathbf{A}_{k,k} = \sum_{m \in \mathcal{M}(k)} \sqrt{\eta_{m,k}^{DL,UC}} \mathbf{L}_k^H \mathbf{H}_{k,m}^H \mathbf{Q}_{k,m}^{DL}$, $\mathbf{R}_k = \sum_{l \neq k} \mathbf{A}_{k,l} \mathbf{A}_{k,l}^H + \sigma_z^2 \mathbf{L}_k^H \mathbf{L}_k$, and $\mathbf{A}_{k,l} = \sum_{m \in \mathcal{M}(k)} \sqrt{\eta_{m,l}^{DL,UC}} \mathbf{L}_k^H \mathbf{H}_{k,m}^H \mathbf{Q}_{l,m}^{DL}$. Letting now

$$\mathbf{B}_{k,k,m} = \mathbf{L}_k^H \mathbf{H}_{k,m}^H \mathbf{Q}_{k,m}^{DL} \Longrightarrow \mathbf{A}_{k,k} = \sum_{m \in \mathcal{M}(k)} \sqrt{\eta_{m,k}^{DL,UC}} \mathbf{B}_{k,k,m}$$

$$\mathbf{B}_{k,l,m} = \mathbf{L}_k^H \mathbf{H}_{k,m}^H \mathbf{Q}_{l,m}^{DL} \Longrightarrow \mathbf{A}_{k,l} = \sum_{m \in \mathcal{M}(k)} \sqrt{\eta_{m,l}^{DL,UC}} \mathbf{B}_{k,l,m}$$

the covariance matrix can be rewritten as

$$\mathbf{R}_k = \sum_{l \neq k} \sum_{m \in \mathcal{M}(l)} \sum_{m' \in \mathcal{M}(l)} \sqrt{\eta_{m,l}^{DL,UC} \eta_{m',l}^{DL,UC}} \mathbf{B}_{k,l,m} \mathbf{B}_{k,l,m'}^H. \quad (11)$$

Using the above notation, the $k$-th user ASE can be expressed as:

$$\mathcal{R}_k = B \log_2 \det[\mathbf{I} + \mathbf{R}_k^{-1} \sum_{m,m'} \sqrt{\eta_{m,k}^{DL,UC} \eta_{m',k}^{DL,UC}} \mathbf{B}_{k,k,m} \mathbf{B}_{k,k,m'}^H]. \quad (12)$$

Using well-known logarithm properties, we also have:

$$\mathcal{R}_k = B \log_2 \underbrace{\left| \sigma_z^2 \mathbf{L}_k^H \mathbf{L}_k + \sum_{l}^{K} \sum_{m} \sum_{m'} \sqrt{\eta_{m,l} \eta_{m',l}} \mathbf{B}_{k,l,m} \mathbf{B}_{k,l,m'}^H \right|}_{g_1(\boldsymbol{\eta})}$$

$$- B \log_2 \underbrace{\left| \sigma_z^2 \mathbf{L}_k^H \mathbf{L}_k + \sum_{l \neq k}^{K} \sum_{m} \sum_{m'} \sqrt{\eta_{m,l} \eta_{m',l}} \mathbf{B}_{k,l,m} \mathbf{B}_{k,l,m'}^H \right|}_{g_2(\boldsymbol{\eta})} \quad (13)$$

## IV. DOWNLINK GEE MAXIMIZATION

Mathematically the problem is formulated as the optimization program:

$$\max_{\boldsymbol{\eta}} \frac{\sum_{k=1}^{K} \mathcal{R}_k(\boldsymbol{\eta})}{\sum_{m=1}^{M} \left[ \sum_{k \in \mathcal{K}(m)} \delta \eta_{m,k} + P_{c,m} \right]} \quad (14a)$$

$$\text{s.t.} \sum_{k \in \mathcal{K}(m)} \eta_{m,k} \leq P_{max,m}, \quad \eta_{m,k} \geq 0. \quad (14b)$$

where $P_{c,m} > 0$ is the circuit power consumed at AP $m$, $\delta \geq 1$ is the inverse of the transmit amplifier efficiency, and $P_{max,m}$ is the maximum transmit power from AP $m$; $\boldsymbol{\eta}$ is a $KM \times 1$ vector containing all the transmit powers of all AP.

[3]The expressions for the CF case follow, as already discussed, as a special case.

Problem (14) is challenging due to its fractional objective, which has a non-concave numerator. This prevents the direct use of standard fractional programming methods such as Dinkelbach's algorithm to solve (14) with affordable complexity. In addition, another issue is represented by the large number of optimization variables, i.e. $KM$. To counter both issues, we resort to the successive lower-bound maximization method. In brief, this method tackles (14) by alternatively optimizing the transmit powers of each AP – we denote by $\boldsymbol{\eta}_m$ the $m$-th AP vector of powers –, while keeping the transmit powers of the other APs fixed. Additionally, each subproblem is tackled by means of sequential optimization. The concave lower bound used at each iteration is expressed as:

$$\mathcal{R}_k = g_1(\boldsymbol{\eta}_m) - g_2(\boldsymbol{\eta}_m)$$
$$\geq g_1(\boldsymbol{\eta}_m) - g_2(\boldsymbol{\eta}_{m,0}) - \nabla_{\boldsymbol{\eta}_m}^T g_2 |_{\boldsymbol{\eta}_{m,0}} (\boldsymbol{\eta} - \boldsymbol{\eta}_{m,0}) \quad (15)$$
$$= \overline{\mathcal{R}}_k(\boldsymbol{\eta} - \boldsymbol{\eta}_{m,0}).$$

The proposed optimization procedure is guaranteed to converge and upon convergence enjoys first-order optimality properties. Further details are omitted due to space constraints.

*1) An alternative definition for the GEE:* The definition of the GEE reported in (14) considers a circuit power consumption that does not depend on the transmit power used by each base station. However, a base station that does not transmit will consume a lower circuit power, since it will switch to idle mode. To account for this circumstance, the terms $P_{c,m}$ can be further detailed to depend on the actual used transmit power, namely defining for all $m = 1, \ldots, M$,

$$P_{c,m} = \widetilde{P}_{c,m} I[P_T(m) > 0] + 0.5(1 - I[P_T(m) > 0]) \quad (16)$$

where $P_T(m) = \sum_k \eta_{m,k}$ is the power radiated by the $m$-th AP and $I[P_T(m) > 0]$ is the indicator function of the set $[P_T(m) > 0]$, being 1 when $P_T(m) > 0$ and 0 otherwise. According to (16), we assume that the circuit power consumption is halved when it does not radiate any power. In this case the derived algorithm is still guaranteed to monotonically increase the GEE value after each iteration, but no first-order optimality can be guaranteed upon convergence, owing to the non-differentiability of (16).

## V. NUMERICAL RESULTS

We consider a carrier frequency of $f_0 = 73$ GHz, a bandwidth of $B = 200$ MHz. and an Open Square scenario of size $250 \times 250$ m$^2$. The additive white noise at the receiver has a power spectral density of $-174$ dBm/Hz and the receiver noise figure is $F = 6$ dB. The simulated system has $M = 100$ APs equipped with $N_{AP} = 16$ antennas each, while the MSs have $N_{MS} = 8$ antennas each; the multiplexing order is $P = 1$. The presented results show the GEE [Mbit/Joule] as defined in (14); the numerical values come from an average over 50 independent channel scenarios as well as users and access points locations. All APs have the same maximum feasible downlink transmit power $P_{max}$. The transmit amplifier efficiency of each transmitter is $\delta = 1$, while the hardware circuit power was modeled according to (16) for each AP, with $\widetilde{P}_{c,m} = 1$ W, for all $m$. Fig. 1 compares the GEE value versus $P_{max}$ achieved by the proposed GEE-maximizing power control scheme (labeled as OPT) and by uniform power allocation (labeled as UNI), considering the UC and CF approaches in the following scenarios: *(i)* perfect CSI and FD beamforming; *(ii)* perfect CSI and HY beamforming, with 4 RF chains used in the *BCD-SD* HY beamforming algorithm; *(iii)* imperfect CSI and FD beamforming, with pilot sequences of length $\tau_p = 64$ and uplink transmit power of $1$ mW; and *(iv)* imperfect CSI and HY beamforming, with pilot sequences of length $\tau_p = 64$ and uplink transmit power of $1$ mW, with 4 RF chains used in the *BCD-SD* HY beamforming algorithm. The results consider a lightly loaded scenario with $K = 5$ MSs and $N = 2$ for the UC case. Inspecting the figures, several considerations can be made. First of all, we see that the proposed power optimization method provides better performance than the uniform power allocation scheme. This is not always true for the case with imperfect CSI, as a consequence of the fact that, in the imperfect CSI case, the optimization was performed based on the estimated channels, since at the design stage the true channel realizations are not available. However, the results shown in Fig. 1 are in terms of the true GEE, i.e. the GEE computed using the true channel realizations, instead of the estimated ones. Next, as expected, we notice that FD beamforming and the availability of perfect CSI lead to better performance. It is also seen that the UC approach generally provides better performance that the plain CF approach, expecially when uniform power control is considered. Fig. 2 considers, instead, a heavily loaded scenario with $K = 20$ MSs, reporting only the UC approach with $N = 1$ and $N = 3$. Also in this case results confirm similar trends as the ones observed in the lightly loaded scenario. For the case $N = 1$ with HY beamfoming and incomplete CSI, the UC ASE is 30 bit/s/Hz, which corresponds to an average rate per user of 300 Mbit/s.

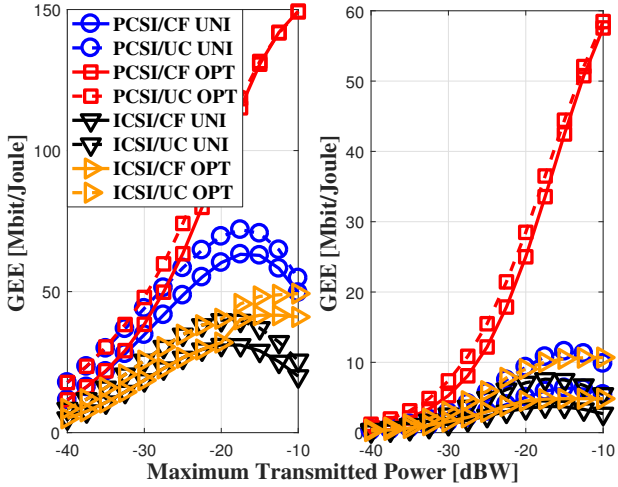

Figure 1. GEE with FD (on the left) and HY beamforming (on the right) versus maximum transmit power. System parameters: $M = 100$, $K = 5$, $N_{AP} \times N_{MS} = 16 \times 8$, $P = 1$, $N = 2$, $\delta = 1$, $P_c = 1$ W.

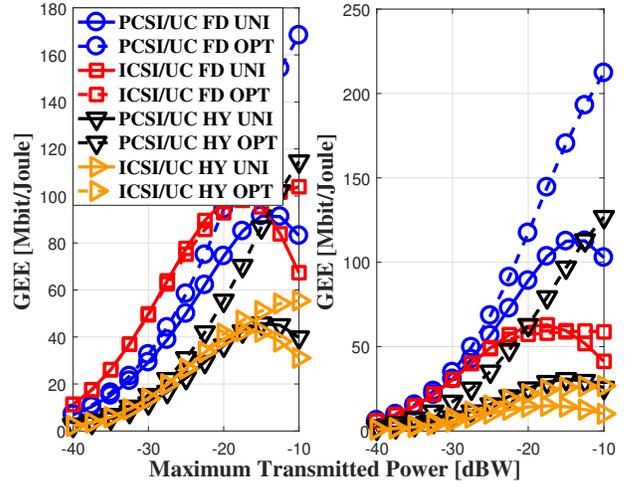

Figure 2. UC approach: GEE with N=1 (on the left) and with N=3 (on the right) versus maximum transmit power. System parameters: $M = 100$, $K = 20$, $N_{AP} \times N_{MS} = 16 \times 8$, $P = 1$, $\delta = 1$, $P_c = 1$ W.

## VI. CONCLUSION

The paper has presented results on the comparison between the CF and UC approaches at mmWave frequencies, with GEE-maximizing power control. The obtained results can be considered as baseline for future work on CF and UC massive MIMO at mmWave.

ACKNOWLEDGEMENT

The paper has been supported by the MIUR program "Dipartimenti di Eccellenza 2018-2022".